\documentclass[aps,prl,preprint,showpacs]{revtex4}
\begin{document}
\title{Black hole hydrodynamics}
\author{Yuri N. Obukhov}
\altaffiliation[On leave from: ]{Dept. of Theoret. Physics, 
Moscow State University, 117234 Moscow, Russia}\email{yo@thp.uni-koeln.de}
\affiliation{Institute for Theoretical Physics, University of Cologne,
50923 K\"oln, Germany}

\begin{abstract}
The curved geometry of a spacetime manifold arises as a solution of Einstein's
gravitational field equation. We show that the metric of a spherically 
symmetric gravitational field configuration can be viewed as an {\it optical 
metric} created by the moving material fluid with nontrivial dielectric and 
magnetic properties. Such a ``hydrodynamical" approach provides a simple 
physical interpretation of a horizon. 
\end{abstract}
\maketitle

\section{Introduction}

Recently, there has been a growing interest in the black hole type
geometries which do not arise due to the genuine gravitational interaction 
but are formed due to the nontrivial (electric and magnetic) properties and 
dynamics of the matter \cite{sound,die,other,nonlin}. When sound or light 
propagates in such a matter, it feels the effective optical metric and moves 
along the corresponding geodesics. In particular, the formation of the horizon
type structures have been demonstrated in many cases. In other words, the 
``effective black holes" were discovered in these non-gravitational physical 
models. 

In this paper we look in the opposite direction. Namely, we consider the
{\it genuine black holes} of the gravitational theory and demonstrate that 
the corresponding metric can be interpreted as the {\it optical metric} 
arising due to the inhomogeneous motion of the material fluid in the
flat Minkowski spacetime. Such an view can be useful in understanding
the characteristic properties of black holes.

In plain words, whereas in \cite{sound,die,other,nonlin} the black hole
is ``fictious" (or effective) arising from the real dynamics of matter,
here we consider the real black hole which corresponds to a ``fictious" 
(or effective) dynamics of matter. Since the latter is a fluid, we can
call such a model ``a hydrodynamics of a black hole". 

\section{Optical geometry}

Let us consider a fluid with nontrivial
permittivity $\varepsilon$ and permeability $\mu$ constants which
moves in a spacetime with the metric $\widetilde{g}_{ij}$. We will be
mainly interested in the case of the Minkowski flat space metric, 
$\widetilde{g}_{ij}=\eta_{ij}$, however for the sake of generality we 
will keep $\widetilde{g}_{ij}$ arbitrary. The motion is described by the 
4-velocity vector field $u^i$ which is normalized as $\widetilde{u}^2
:= u^iu^j \widetilde{g}_{ij} = c^2$. 

In such a medium, the light is propagating along the null geodesics
not of the original spacetime metric $\eta_{\mu\nu}$ but of the 
so called optical metric, see \cite{Gordon,Book},
\begin{equation}
g_{ij} = \widetilde{g}_{ij} + (1/n^2 - 1)\,u_i u_j/c^2,\label{gmn1}
\end{equation}
where $n = \sqrt{\varepsilon\mu}$ is the refraction index of the fluid.
The inverse metric reads
\begin{equation}
g^{ij} = \widetilde{g}^{ij} + (n^2 - 1)\,u^i u^j/c^2.\label{gmn2}
\end{equation}
It is worthwhile to note that the velocity has a different normalization
for the optical metric: $u^2 := u^iu^j g_{ij} = c^2/n^2$ which is evident
from (\ref{gmn1}).

The kinematic properties of the fluid flow with respect to the metric
$g_{ij}$ are described by the tensors of shear $\sigma_{ij}$, vorticity 
$\omega_{ij}$, the volume expansion $\theta$, and the acceleration $a^i$. 
These objects are constructed from the components of the covariant 
derivative $\nabla_k u^i$ of the 4-velocity with the help of the projector 
$h^i_j = \delta^i_j - u^ig_{jk}u^k/u^2$ as follows: $\sigma_{ij} := h^k{}_i
h^l{}_j \nabla_{(k}u^m\,g_{l)m} - {\frac 1 3}\,g_{ik}\,h^k_j\nabla_lu^l$, 
$\omega_{ij} := h^k{}_ih^l{}_j \nabla_{[k} u^m\,g_{l]m}$, 
$\theta := \nabla_lu^l$, $a^i := u^k\nabla_k u^i$. If we replace in
these formulas $g\rightarrow\widetilde{g}$ (including the replacement of
the covariant derivatives $\nabla\rightarrow\widetilde{\nabla}$), we
find the kinematics of the fluid with respect to the background 
spacetime metric. The latter quantities will be denoted by the tildes. 
It is straightforward to find the relation between the two sets of the
kinematic objects, $\omega, \sigma, \theta, a$, and $\widetilde{\omega}, 
\widetilde{\sigma}, \widetilde{\theta}, \widetilde{a}$. 

At first, we notice that the Riemannian connection (Christoffel symbols) 
for the optical metric (\ref{gmn1})-(\ref{gmn2}) reads: 
\begin{eqnarray}
\Gamma^i_{jk} &=& \widetilde{\Gamma}^i_{jk} + (1 - n^2)\Big( a^i\,u_ju_k
+ {\frac 13}\,\theta\,(\widetilde{g}_{jk} - u_ju_k)\,u^i \nonumber\\
&& \qquad +\,u^i\,\sigma_{jk} + u_j\,\omega_k{}^i + u_k\,\omega_j{}^i
\Big).\label{conn}
\end{eqnarray}
Note that the indices on the right-hand side are lowered and raised with
the help of the background spacetime metric $\widetilde{g}$. 
Now it is easy to verify that
\begin{equation}\label{relat}
\widetilde{\sigma}_{ij} = \sigma_{ij},\quad \widetilde{\theta}=\theta,\quad 
\widetilde{\omega}_{ij} = n^2\,\omega_{ij},\quad \widetilde{a}^i = n^2\,a^i. 
\end{equation}
In other words, the kinematic properties of the fluid are basically the 
same as seen with respect to the spacetime metric $\widetilde{g}$ or with 
respect to the optical metric $g$. When the fluid is non-refractive, i.e. 
$n=1$, the two metrics, as well as the connection (\ref{conn}) and 
the kinematic objects, coincide. 

\section{Spherically symmetric configurations}

Let us consider the Minkowski spacetime with the metric 
$d\widetilde{s}^2 = \widetilde{g}_{ij}\,dx^idx^j = c^2dt^2 - dr^2 
- r^2d\theta^2 - r^2\sin^2\theta\,d\varphi^2$ in the spherical 
coordinate system $(t, r, \theta, \varphi)$. Let the radial flow of the 
fluid be described by the 4-velocity $u^i = \gamma\,(1, c\beta, 0, 0)$ with
\begin{equation}
\beta = \pm\sqrt{\frac {1 - f(r)}{n^2 - f(r)}}\,,\qquad 
\gamma = \sqrt{\frac {n^2 - f(r)}{n^2 - 1}}\,.\label{bega}
\end{equation}
Then (\ref{gmn2}) yields the components of the optical metric 
\begin{eqnarray}
g^{00} &=& {\frac 1 {c^2}}\,(n^2 + 1 - f),\label{g00}\\
g^{01} &=& {\frac \beta c}\,(n^2 - f),\label{g01}\\
g^{11} &=& -\,f,\label{g11}\\
g^{22} &=& \sin^2\theta \, g^{33}\, = \,-\,\frac{1}{r^2} .
\label{g33}
\end{eqnarray}
The corresponding optical line element reads
\begin{eqnarray}
ds^2 &=& g_{ij}\,dx^idx^j = f\,{\frac {c^2}{n^2}}\,dt^2 
+ 2\,{\frac {c\beta}{n^2}}\,(n^2 - f)\,dtdr\nonumber\\ 
&& + {\frac {(f - n^2 - 1)}{n^2}} dr^2 
- r^2d\theta^2 - r^2\sin^2\theta d\varphi^2.\label{ds2opt}
\end{eqnarray}
For an arbitrary refraction index $n$ and for any function $f=f(r)$
this interval describes the spherically symmetric static geometry
which can be recasted into the simple Schwarzschild form
\begin{equation}
ds^2 = f\,c^2d{t'}^2 - {\frac 1 f}\,dr^2 
- r^2d\theta^2 - r^2\sin^2\theta\,d\varphi^2,
\end{equation}
if we change the original time coordinate to the new one:
\begin{equation}
t' = {\frac 1 n}\left( t \mp \int dr\,{\frac 
{\sqrt{(1-f)(n^2 - f)}} {cf}}\,\right).\label{tnew1}
\end{equation}

Let us study the motion of light in the optical geometry (\ref{ds2opt}).
The light propagates along the null geodesics with the tangent vectors 
$k^i=\dot x^\mu=(\dot{t},\dot{r},\dot{\theta}, \dot{\varphi})$ in such a way 
that $g_{ij}k^ik^j =0$. The most interesting case is represented by the 
purely radial propagation of a photon which corresponds to $k^2=k^3=0$. As a 
result, the null condition $g_{00}(k^0)^2 + 2g_{01}k^0k^1 + g_{11}(k^1)^2 =0$ 
yields the equation for the radial path:
\begin{equation}
\dot{t} = {\frac {-\beta (n^2 - f) \pm n}{cf}}\,\dot{r}.\label{dotrt}
\end{equation}
The plus (minus) sign corresponds to the ray propagating from (to) the 
origin. Now let us take into account that the function $f$ is directly 
related to the velocity of the fluid, see (\ref{bega}). From the latter 
equation we find $f = (1 - n^2\beta^2)/(1 - \beta^2)$. Inserting this
into (\ref{dotrt}), we obtain that the radial coordinate of the photon 
changes in time as
\begin{equation}
{\frac{\ dr}{cdt}} = {\frac {n\beta\pm 1}{n\pm \beta}}
\label{drdt}.
\end{equation}
When the fluid flows {\it outwards}, i.e., $\beta>0$, the {\it incoming} 
light (with $\dot{r}<0$) evidently ``freezes'', with the vanishing of 
$dr/dt$, at the radius $r_h$ defined $n\beta\vert_{r_h}=1$. The same takes 
place for an {\it outgoing} light in the {\it inwardly} directed flow,
when the light becomes ``freezed'' at $n\beta\vert_{r_h}= - 1$. Both cases
are covered by the condition
\begin{equation}\label{hor}
n^2\beta^2\,\vline\,{\hbox{\raisebox{-1.5ex}{\scriptsize{$r_h$}}}} = 1.
\end{equation}
In all cases, the ``freezing'' is absent for the light propagating in the 
{\it same} direction as the moving dielectric medium.

There is a clear physical interpretation of the above result \cite{bhdie}. 
Recall that $\beta = v/c$ for the 3-velocity $v$ of the matter flow, and 
$v_c = c/n$ is the velocity of light in a medium with refraction index $n$ 
in the absence of external fields. Using this, we can recast the equation 
(\ref{drdt}) into the form of the relativistic transformation of the velocity:
\begin{equation}
\frac{dr}{dt}= {\frac {v\pm v_c}{1\pm vv_c/c^2}}.\label{drdt1}
\end{equation}
In other words, the propagation rate of light with respect to the 
flat Minkowski spacetime (the laboratory reference system) is determined
as a relativistic sum of the velocity of light in the medium $v_c$ and
the relative velocity $v$ of the fluid (the moving reference frame)
The ``freezing" of light takes place at a surface on which the velocity
of matter $v$ becomes equal to the velocity of light $v_c$ in the medium:
the light cannot ``overcome" the flow after $r = r_h$. 

The above observation thus gives a transparent physical interpretation 
of the horizon. Below we give two illustrative examples.

\subsection{Schwarzschild black hole}

As a first example we take the Schwarzschild black hole. It is described by 
\begin{equation}
f(r) = 1 - {\frac {r_0} r},\qquad r_0 = 2Gm/c^2.\label{fs}
\end{equation}
The coordinate transformation (\ref{tnew1}) then reads explicitly
\begin{eqnarray}
t' &=& {\frac 1 n}\Bigg[t\mp {\frac {r_0} c}\Bigg(2\sqrt{(n^2 - 1)r/r_0 + 1}
\nonumber\\ 
&& \quad +\,n\,\ln{\frac {\sqrt{(n^2 - 1)r/r_0 + 1} - n}
{\sqrt{(n^2 - 1)r/r_0 + 1} + n}}\Bigg)\Bigg].\label{tnew2}
\end{eqnarray}
Note that (\ref{tnew2}) does not cover the case $n=1$, when one finds
$t' = t \mp {\frac {r_0} c}\,\log(r/r_0 - 1)$ instead. 

As it is clear from (\ref{bega}), the condition (\ref{hor}) of ``freezing" 
is fulfilled when $f=0$. Geometrically, this corresponds to the Schwarzschild 
horizon, $r_h = r_0$, see (\ref{fs}). 

\subsection{De Sitter geometry}

As a second example, we take the de Sitter spacetime. As it is well known,
the de Sitter world has many faces in different coordinate systems. Its 
static spherically symmetric realization arises for 
\begin{equation}
f = 1 - {\frac \Lambda 3}\,r^2.\label{desitter}
\end{equation}
Here $\Lambda$ is the cosmological constant which determines the de Sitter
radius. Actually, this is only a particular case of the more general metric
with $f = 1 - r_0/r - \Lambda r^2/3$ (Schwarzschild-de Sitter or 
Kottler metric), but we on purpose limit our attention to the 
spacetime of constant curvature determined by (\ref{desitter}). 

The corresponding coordinate transformation (\ref{tnew1}) is then given 
explicitly by
\begin{eqnarray}
t' &=& {\frac 1 n}\Bigg[t\pm {\frac {1} {2c}}\sqrt{\frac 3\Lambda}\Bigg(
2\sqrt{n^2 - 1 + \Lambda r^2/3} \nonumber\\ 
&& \quad +\,n\,\ln{\frac {\sqrt{n^2 - 1 + \Lambda r^2/3} - n}
{\sqrt{n^2 - 1 + \Lambda r^2/3} + n}}\Bigg)\Bigg].\label{tnew3}
\end{eqnarray}

In this case, the cosmological horizon arises at $r_h = \sqrt{3/\Lambda}$.

\section{Discussion}

Hydrodynamical interpretation of the spherically symmetric gravitational
field appears to be useful for understanding the physical properties of 
such configurations. In particular, both the black hole horizons and the 
cosmological horizons are obtained as the surfaces on which the flow of
the (fictious) fluid overpowers the propagation of light. In this sense,
gravity becomes encoded into the dynamics of the matter in the flat 
Minkowski spacetime. 

The result obtained confirms the conclusions of the physical models
considered in \cite{sound,die,other,nonlin,bhdie}. 

It is interesting to note that in the Schwarzschild case, the fluid is
at rest ($\beta = 0$) at the spatial infinity and its velocity reaches
the speed of light ($\beta = 1$) at the origin. The opposite case is
represented by the de Sitter model where $\beta = 0$ is at the origin
and $\beta = 1$ at the spatial infinity. One can consider a more nontrivial
geometry of Reissner-Nordstr\"om-de Sitter with $f = 1 - r_0/r + e^2/r^2 
- \Lambda r^2/3$. Here the two (the black and the cosmological) horizons are
present, and the corresponding hydrodynamical model describes a flow which
reaches the velocity of light both at the origin and at the spatial 
infinity. 

A generalization of the above results to a rotating black hole is 
nontrivial. The fluid in this case should move not only radially but
with an additional axial twist. 

It may be noticed that the coordinate transformation (\ref{tnew1}),
(\ref{tnew2}), (\ref{tnew3}) is very much analogous to the other regular 
transformations on the horizon, see \cite{DF,wil,martel}, for example. 
This point is emphasized in the recent paper \cite{rosq} which essentially
overlaps with our observations.

\section{Acknowledgments}

This work was supported by the Deutsche Forschungsgemeinschaft (Bonn) 
with the grant HE~528/20-1.


\begin{thebibliography}{88}

\bibitem{sound}
W.G. Unruh, {\it Experimental Black-Hole Evaporation?},
{\sl Phys. Rev. Lett.} {\bf 46} (1981) 1351-1353;
W.G. Unruh, {\it Sonic analogue of black holes and the effects of high 
frequencies on black hole evaporation}, {\sl Phys. Rev.} {\bf D51} (1995) 
2827-2838;
L.J. Garay, J.R. Anglin, J.I. Cirac, and P. Zoller,
{\it Sonic black holes in dilute Bose-Einstein condensates},
{\sl Phys. Rev.} {\bf A63} (2001) 023611 (13 pages);
M. Visser, {\it Acoustic black holes: horizons, ergospheres and Hawking 
radiation}, {\sl Class. Quantum Grav.} {\bf 15} (1998) 1767-1791.

\bibitem{die}
U. Leonhardt, {\it Space-time geometry of quantum dielectrics},
{\sl Phys. Rev.} {\bf A62} (2000) 012111 (8 pages);
U. Leonhardt and P. Piwnicki, {\it Optics of nonuniformly moving media},
{\sl Phys. Rev.} {\bf A60} (1999) 4301-4312;
I. Brevik and G. Halnes, {\it Light rays at optical black holes in 
moving media}, {\sl Phys. Rev.} {\bf D65} (2002) 024005 (12 pages);
R. Sch\"utzhold, G. Plunien, and G. Soff, {\it Dielectric black hole analogs},
{\sl Phys. Rev. Lett.} {\bf 88} (2002) 061101 (4 pages);
S. Liberati, S. Sonego, and M. Visser, {\it Scharnhorst effect at oblique 
incidence}, {\sl Phys. Rev.} {\bf  D63} (2001) 085003 (10 pages). 

\bibitem{other} 
B. Reznik, {\it Origin of the thermal radiation in a solid-state analogue 
of a black hole}, {\sl Phys. Rev.} {\bf D62} (2002) 044044 (7 pages);
R. Sch\"utzhold and W.G. Unruh, {\it Gravity wave analogues of black holes},
{\sl Phys. Rev.} {\bf D66} (2002) 044019 (13 pages).

\bibitem{nonlin}
V.A. De Lorenci and R. Klippert,
{\it Analogue gravity from electrodynamics in nonlinear media},
{\sl Phys. Rev.} {\bf D65} (2002) 064027 (6 pages);
V.A. De Lorenci and M.A. Souza, {\it Electromagnetic wave propagation 
inside a material medium: an effective geometry interpretation}, 
{\sl Phys. Lett.} {\bf B512} (2001) 417-422;
Yu.N. Obukhov and G.F. Rubilar, {\it Fresnel analysis of wave propagation 
in nonlinear electrodynamics}, {\sl Phys. Rev.} {\bf D66} (2002) 024042 
(11 pages).

\bibitem{Gordon} 
W. Gordon, {\em Zur Lichtfortpflanzung nach der Relativit\"atstheorie}, 
{\sl Ann. der Phys.} {\bf 4} (1923) 421-456.

\bibitem{Book} 
F.W. Hehl and Yu.N. Obukhov, {\it Foundations of Classical Electrodynamics:
Charge, Flux, and Metric} (Birkh\"auser: Boston, MA, 2003).

\bibitem{bhdie}
V.A. De Lorenci, R. Klippert, and Yu.N. Obukhov, {\it Optical black holes
in moving dielectrics}, {\sl Phys. Rev.} {\bf D68} (2003) in print; 
gr-qc/0210104. 

\bibitem{DF}
D. Finkelstein, {\it Past-future asymmetry of the gravitational field 
of a point particle}, {\sl Phys. Rev.} {\bf 110} (1958) 965-967.  

\bibitem{wil}
M.K. Parikh and F. Wilczek, {\it Hawking radiation as tunneling},
{\sl Phys. Rev. Lett.} {\bf 85} (2000) 5042-5045.

\bibitem{martel}
K. Martel and E. Poisson, {\it Regular coordinate systems for Schwarzschild
and other spherical spacetimes}, {\sl Am. J. Phys.} {\bf 69} (2001) 476-480.

\bibitem{rosq}
K. Rosquist, {\it A moving medium simulation of Schwarzschild black hole 
optics}, 4 pages. E-print: gr-qc/0309104. 

\end{thebibliography}
\end{document}